\documentstyle[12pt,aaspp4]{article}

\begin{document}

%
\def\n{\footnotemark}
\def\IUE{{\it IUE}}
\def\HST{{\it HST}}
\def\ISO{{\it ISO}}
\def\deg{$^{\rm o}$}
\def\degC{$^{\rm o}$C}
\def\arcsec{\ifmmode '' \else $''$\fi}
\def\arcmin{$'$}
\def\arcsecpoint{\ifmmode ''\!. \else $''\!.$\fi}
\def\arcminpoint{$'\!.$}
\def\kms{\ifmmode {\rm km\ s}^{-1} \else km s$^{-1}$\fi}
\def\Msun{\ifmmode {\rm M}_{\odot} \else M$_{\odot}$\fi}
\def\Lsun{\ifmmode {\rm L}_{\odot} \else L$_{\odot}$\fi}
\def\Zsun{\ifmmode {\rm Z}_{\odot} \else Z$_{\odot}$\fi}
\def\ergsAcm{ergs\,s$^{-1}$\,cm$^{-2}$\,\AA$^{-1}$}
\def\ergscm2{ergs\,s$^{-1}$\,cm$^{-2}$}
\def\qo{\ifmmode q_{\rm o} \else $q_{\rm o}$\fi}
\def\Ho{\ifmmode H_{\rm o} \else $H_{\rm o}$\fi}
\def\ho{\ifmmode h_{\rm o} \else $h_{\rm o}$\fi}
\def\ltsim{\raisebox{-.5ex}{$\;\stackrel{<}{\sim}\;$}}
\def\gtsim{\raisebox{-.5ex}{$\;\stackrel{>}{\sim}\;$}}
\def\vFWHM{\ifmmode v_{\mbox{\tiny FWHM}} \else
            $v_{\mbox{\tiny FWHM}}$\fi}
\def\CCF{\ifmmode F_{\it CCF} \else $F_{\it CCF}$\fi}
\def\ACF{\ifmmode F_{\it ACF} \else $F_{\it ACF}$\fi}
\def\Halpha{\ifmmode {\rm H}\alpha \else H$\alpha$\fi}
\def\Hbeta{\ifmmode {\rm H}\beta \else H$\beta$\fi}
\def\Hgamma{\ifmmode {\rm H}\gamma \else H$\gamma$\fi}
\def\Hdelta{\ifmmode {\rm H}\delta \else H$\delta$\fi}
\def\Lya{\ifmmode {\rm Ly}\alpha \else Ly$\alpha$\fi}
\def\Lyb{\ifmmode {\rm Ly}\beta \else Ly$\beta$\fi}
\def\Lyg{\ifmmode {\rm Ly}\beta \else Ly$\gamma$\fi}
\def\hi{H\,{\sc i}}
\def\hii{H\,{\sc ii}}
\def\hei{He\,{\sc i}}
\def\heii{He\,{\sc ii}}
\def\ci{C\,{\sc i}}
\def\cii{C\,{\sc ii}}
\def\ciii{\ifmmode {\rm C}\,{\sc iii} \else C\,{\sc iii}\fi}
\def\civ{\ifmmode {\rm C}\,{\sc iv} \else C\,{\sc iv}\fi}
\def\ni{N\,{\sc i}}
\def\nii{N\,{\sc ii}}
\def\niii{N\,{\sc iii}}
\def\niv{N\,{\sc iv}}
\def\nv{N\,{\sc v}}
\def\oi{O\,{\sc i}}
\def\oii{O\,{\sc ii}}
\def\oiii{O\,{\sc iii}}
\def\o5007{[O\,{\sc iii}]\,$\lambda5007$}
\def\oiv{O\,{\sc iv}}
\def\ov{O\,{\sc v}}
\def\ovi{O\,{\sc vi}}
\def\neiii{Ne\,{\sc iii}}
\def\nev{Ne\,{\sc v}}
\def\neviii{Ne\,{\sc viii}}
\def\mgi{Mg\,{\sc i}}
\def\mgii{Mg\,{\sc ii}}
\def\mgx{Mg\,{\sc x}}
\def\siIV{Si\,{\sc iv}}
\def\siIII{Si\,{\sc iii}}
\def\siII{Si\,{\sc ii}}
\def\siI{Si\,{\sc i}}
\def\si{S\,{\sc i}}
\def\sii{S\,{\sc ii}}
\def\siii{S\,{\sc iii}}
\def\siv{S\,{\sc iv}}
\def\sv{S\,{\sc v}}
\def\svi{S\,{\sc vi}}
\def\caii{Ca\,{\sc ii}}
\def\fei{Fe\,{\sc i}}
\def\feii{Fe\,{\sc ii}}
\def\feiii{Fe\,{\sc iii}}
\def\alii{Al\,{\sc ii}}
\def\aliii{Al\,{\sc iii}}
\def\piv{P\,{\sc iv}}
\def\pv{P\,{\sc v}}
\def\cliv{Cl\,{\sc iv}}
\def\clv{Cl\,{\sc v}}
\def\nai{Na\,{\sc i}}
\def\o{\o}
%

\title{The UV--Optical Albedo of Broad Emission Line Clouds}

\author{Kirk Korista}
\affil{Department of Physics \& Astronomy, University of Kentucky,
Lexington, KY 40506 and Department of Physics, Western Michigan University, Kalamazoo, MI 49008}
\author{Gary Ferland}
\affil{Department of Physics \& Astronomy, University of Kentucky,
Lexington, KY 40506}

\begin{abstract}
{We explore the effective UV--optical albedos of a variety of types of
broad emission line clouds, as well as their possible effects on the
observed spectra of AGN. An important albedo source in moderately
ionized ionization-bounded clouds is that due to neutral hydrogen:
Rayleigh scattering of continuum photons off the extreme damping wings
of \Lya\/. The photons resulting from this scattering mechanism may
contribute significantly to the \Lya\/ emission line, especially in the
very broad wings. In addition, line photons emitted near 1200~\AA\/
(e.g., \nv\/ $\lambda$1240) that stream toward the neutral portion of
the cloud may be reflected off this Rayleigh scattering mirror, so that
they preferentially escape from the illuminated face. Inclusion of this
effect can alter predicted emission line strengths and profiles. In
more highly-ionized ionization-bounded clouds, Thompson scattering
dominates the UV--optical albedo, but this albedo is lessened by the
hydrogen gas opacity --- these clouds are most reflective on the long
wavelength side of the hydrogen recombination edges. This feature may
then alter the shapes of the spectral regions near the recombination
edges, e.g., the Balmer jump.  We illustrate the effects of gas density
and line broadening on the effective albedo.  We also discuss the
reflection effects of the accretion disk and the ``dusty torus.'' The
accretion disk is an effective reflector of UV--optical photons,
whether by electron or Rayleigh scattering, and it is possible that we
observe a significant fraction of this light from the AGN in
reflection.  This effect can alter the emission line profiles and even
destroy emission at the Lyman jump emitted by broad line clouds.
Finally, we discuss the possibility that continuum reflection from
broad line clouds is at least in part responsible for the polarized
broad absorption line troughs.}

\end{abstract}

\keywords{galaxies: Seyfert --- quasars: general --- scattering}
\newpage
%
%
%
\section{Introduction}

The presence of ``mirrors'' in the vicinity of Active Galactic Nuclei
(AGN) has been known for more than a decade, mainly via optical/UV
spectropolarimetric observations as well as X-ray observations. Along
some lines of sight, the light from the central continuum source and
the broad emission line clouds is blocked from our direct view by the
putative molecular torus and reflected by material (warm electrons and
dust) lying outside the torus opening. The nuclear emission is observed
in polarized light in many Seyfert 2 galaxies (Miller \& Antonucci
1983; Antonucci \& Miller 1985; Miller \& Goodrich 1990; Tran, Miller,
\& Kay 1992; Goodrich \& Miller 1994). The ``reflection bump '' or
``Compton mirror'' is the reflection of hard X-rays off some object
which covers $\sim$~2$\pi$ steradians of the sky as seen from the hard
X-ray source (Lightman \& White 1988; Ross \& Fabian 1993; Nandra \&
George 1994; \.{Z}ycki et al.\ 1994; Poutanen et al.\ 1996). The
accompanying Fe K$\alpha$ emission line requires high column density
gas and is thought to arise in the reflecting gas. This mirror is often
taken to be the accretion disk, though Krolik, Madau, \& \.{Z}ycki
(1994) have also proposed the molecular torus.  Observations of broad
absorption line QSOs (BAL QSOs) show the troughs to be more polarized
than the continuum. It has been inferred that part of the rest frame UV
continuum is observed in reflection and/or is partially screened from
view (Glenn, Schmidt, \& Foltz 1994; Cohen et al.\ 1995; Goodrich \&
Miller 1995; Hines \& Wills 1995; Goodrich 1996). Finally, strongly
rising polarization shortward of the expected intrinsic Lyman limits in
2 of the 3 QSOs observed by Koratkar et al.\ (1995) led them to propose
the reflection of an intrinsic Lyman edge in {\em emission}.

Here we examine the UV--optical albedo, $A_{eff}(\lambda) \equiv
F_{refl}(\lambda)/F_{inci}(\lambda)$, of broad emission line clouds
with hydrogen ionization fronts (henceforth, ionization-bounded clouds)
to demonstrate its possible significance to the observed spectra of
AGN. Here, $F_{inci}(\lambda)$ is the incident AGN continuum flux at
the illuminated face of the cloud; $F_{refl}(\lambda)$ is the reflected
flux that includes only the incident continuum photons re-emitted by
the illuminated face of the cloud into $2\pi$ steradians  --- i.e., it
does not include diffuse continua or line emission. A cartoon of the
general situation is shown in Figure~1. 

In $\S$~2 we present the calculations of the effective albedo, and in
$\S$~3 discuss the implications of the albedo of the broad line region
(BLR) clouds to the observed spectra of AGN. Ionization-bounded broad
line clouds have a high albedo near 1200~\AA\/, due to \Lya\/ Rayleigh
scattering of continuum photons. This reflection could account for a
small but significant fraction of the measured \Lya\/ $\lambda$1216
equivalent width. This mirror could also alter the transport of those
lines emitted near \Lya\/ (e.g., \nv\/ $\lambda$1240). We comment
briefly on some of the implications of the existence of Rayleigh and
electron scattering mirrors in the accretion disk. We will also discuss
the possible contribution of the broad line clouds' albedo to the
polarized light observed in the troughs of BAL QSOs. For comparison, we
will illustrate the albedo of a dusty cloud whose parameters roughly
approximate that of the ``molecular torus.''

\section{Calculations}

\subsection{The Effective Albedo: Scattering vs.\ Absorption}

The effective albedo of a cloud is the result of a competition between
scattering and absorption. The dominant scattering mechanism in the UV
in high column density, high ionization broad line clouds is Thomson
(electron) scattering, and that in clouds with significant neutral
hydrogen column densities is Rayleigh scattering off the extreme wings
of atomic hydrogen lines. In the latter case, photons incident upon a
neutral hydrogen slab are scattered monochromatically off the extreme
damping wings of (mainly) \Lya\/ $\lambda$1216, with a cross-section
for interaction similar to that which makes the earth's day sky blue
(Gavrila 1967; Mihalas 1978).  The pure scattering albedo is 1 near
1200~\AA\/ for neutral hydrogen column densities exceeding $\sim
10^{20}$~cm$^{-2}$. We shall associate the Rayleigh scattering feature
with the wavelength 1200~\AA\/ to distinguish it from the emission line
\Lya\/ $\lambda$1216.

The dominant sources of absorption opacity which most affect the
UV--optical albedo of the broad line clouds are bound-free transitions
of the neutral and singly ionized species generally residing behind the
hydrogen ionization front. Foremost among these is atomic hydrogen
opacity, in the various continua (Lyman, Balmer, Paschen, etc).
Photoionization from excited states of He and He$^+$ is also
important.  Next in importance are the ionization edges of the more
abundant atoms and singly ionized species of C, N, O, Mg, Si, S, Ca,
Fe, ionized from the ground and in some cases the excited states. At
large enough neutral hydrogen column densities, the opacities of the
atoms of even the less abundant elements (e.g., Na, Al, P, K, Cr, Mn,
Co, Ni) become significant. (The lightest 30 elements are included in
the calculations; here we assume solar abundances of these elements
[Grevesse \& Anders 1989; Grevesse \& Noels 1993].)

\subsection{Reflection from Broad Line Clouds}

For a cloud of total column density $N(H)$ and density $n(H)$, the
ionization parameter, $U(H) \equiv \Phi(H)/n(H)c$, mainly sets (1) the
ionized hydrogen column density $N(H^+)$ and thus the electron
scattering albedo, (2) the neutral hydrogen column density $N(H^o)$ and
thus the Rayleigh scattering albedo and the hydrogen opacity, and (3)
the opacities of the neutral and singly ionized metals. Thus $N(H)$ and
$U(H)$, for a given incident continuum shape and chemical abundances,
will determine the effective albedo of a BLR cloud. We assume a total
hydrogen particle density of $10^{11}$~cm$^{-3}$, an incident
optical--EUV continuum of the form $f_{\nu} \propto \nu^{-1.2}
exp(-E/415~\rm{eV})$) with an X-ray power law of the form $f_{\nu}
\propto \nu^{-0.9}$ spanning 13.6~eV to 100~keV, and an $\alpha_{ox} =
1.2$.  This SED may be appropriate for some Seyfert~1 galaxies (Walter
et al.\ 1994). The choice of continuum shape does not strongly
influence the general results presented here; however, different
chemical abundances would have a significant impact on the effective
albedo. The impact of gas density will be discussed.  We used the
spectral synthesis code {\sc Cloudy}, version 90.04 (Ferland 1996)
assuming constant density, plane-parallel slabs.

In Figures~2a and 2b we plot the effective UV--optical (800~\AA\/ --
1.0~$\mu$m) albedo ($A_{eff}(\lambda)$) of BLR clouds for a range of
cloud parameters, $U(H)$ and $N(H^o)$, thought typical for the BLR. The
effects of the hydrogen opacity and, in some cases, the metal opacity
are clearly visible.  Unless the Thomson optical depth is significant,
$A_{eff}(\lambda)$ is small except in the vicinity of 1200~\AA\/ and
vanishes below the Lyman limit at 912~\AA\/.  The strength and breadth
of the \Lya\/ Rayleigh scattering feature is striking. We now discuss
some of the important features of Figure~2.

The effective albedo for 5 neutral hydrogen column density clouds
($10^{20} - 10^{25}$~cm$^{-2}$) with $\log U(H) = -2$ are plotted as
light solid lines in Figure~2a. Note the increasing width and strength
of the Rayleigh scattering feature with increasing $N(H^o)$. For all
but the smallest $N(H^o)$ considered ($10^{20}$~cm$^{-2}$), the
$A_{eff}(\lambda)$ is nearly 1 at 1200~\AA\/. The electron scattering
optical depth increases from essentially zero to 0.196, from small to
large column density clouds.  However, note that the corresponding
albedo is never realized outside the reaches of the \Lya\/ Rayleigh
scattering profile.  The reason for this is the gas bound-free opacity
that also increases with column density.  Under these conditions of
relatively low ionization, the effective albedo is significant mainly
in the spectral regions dominated by the Rayleigh scattering.
Increasing the neutral column density beyond $10^{25}$~cm$^{-2}$ does
not increase the effective albedo significantly, since at these large
neutral column densities the gas absorption opacity completely
dominates over the scattering.

While most calculations of broad line clouds assume thermal velocity
fields, including these thus far, significant microturbulent or flow
velocities may be present. If present, the effective albedo of the
cloud can be altered by changes in the gas opacity --- lower optical
depths for \Lya\/ and the excited state transitions of hydrogen (and
helium) will reduce line trapping and hence the importance of
photoionization from the excited states of hydrogen (and helium). Thus,
the Thomson scattering optical depth will be diminished, because of the
reduced ionized hydrogen column density.  For the same reason the
Balmer continuum optical depth will also be reduced, as will optical
depths of other excited state continua of H and He. Furthermore, the
reduced excited state H, He opacities are compensated by increased
photoionization of $\rm{C^o}$ and neutral third and fourth row
elements.  Thus for sufficient column densities (i.e., $\log N(H^o)
\gtsim 24$), the effective albedo of the Rayleigh scattering feature
can increase dramatically in the presence of microturbulence, or in the
presence of Sobolev velocity gradients, for reasons given above.
Compare the bold solid line in Figure~2a, with a microturbulent
velocity of $v_{turb} = 100$~km/s, with its counterpart from thermal
width gas.

At larger ionization parameters, the gas becomes more ionized and the
importance of Thomson scattering to the albedo increases.
$A_{eff}(\lambda)$ is plotted as a dotted line in Figure~2b for $\log
N(H^o) = 23$ and an ionization parameter of $\log U(H) = -1$.  This
curve should be compared to those for $\log U(H) = -2$ and $\log N(H^o)
= 23$ and $\log N(H^o) = 25$, the third and first largest
$A_{eff}(\lambda)$ plotted as solid lines in Figure~2a (with $v_{turb}
= 0$). In the first case the two clouds have the same $N(H^o)$, but
different ionization parameters. The cloud with the larger ionization
parameter has a larger Thomson optical depth, and thus a larger albedo
across most of the spectrum. In the second case the clouds have nearly
the same Thomson optical depth (0.201 $vs$.\ 0.196), however, one of
the clouds has $100 \times$ the neutral hydrogen column density and
thus a much stronger (i.e., broader) Rayleigh scattering feature. The
larger $N(H^o)$ cloud also has a much larger gas opacity and thus its
effective albedo falls below that of the lower column density, higher
$U(H)$ cloud for wavelengths longer than $\sim$~1860~\AA\/.

The light dashed line in Figure~2b is $A_{eff}(\lambda)$ for $\log
N(H^o) = 23$ and an ionization parameter of $\log U(H) = 0$, and should
be compared to the effective albedos of the other clouds with $\log
N(H^o) = 23$, but lower $U(H)$. A cloud with large $U(H)$ and column
density will have a significant Thomson scattering optical depth, here
$\tau_{Th} = 1.47$.  Note the small but significant $A_{eff}(\lambda)$
at wavelengths just longward of the Lyman limit. Even near 1200~\AA\/
about 75\% of $A_{eff}(\lambda)$ is due to electron scattering, unlike
the lower ionization clouds. However, the typical UV--optical
$A_{eff}(\lambda)$ outside the realm of the Rayleigh scattering feature
is still less than the Thomson scattering albedo mainly because of the
hydrogen absorption opacity.  Note, too, that $A_{eff}(\lambda\/1200)$
is smaller (0.89) than for the previous clouds, since the opacities in
the hydrogen Balmer and \hei\/ $2~^3S$ continua are significant enough
to reduce the albedo here.

A broad range in gas densities are likely present in the BLR (Baldwin
et al.\ 1995). Up to now we have considered clouds with $n(H) =
10^{11}~\rm{cm}^{-3}$. At lower densities, e.g., $n(H) =
10^9~\rm{cm}^{-3}$, the flux in radiation at the illuminated face of
the cloud is smaller at the same ionization parameter, and thus the
atoms and ions have smaller excited state populations. In the case of
hydrogen (the major electron donor and opacity source), this has three
effects: (1) photoionization from excited states and collisional
ionization becomes less important and the gas is less ionized, lowering
the Thomson scattering albedo; (2) the recombination continua optical
depths (e.g., Balmer) become smaller, so the drop in effective albedo
shortward of the continuum edges is less pronounced; (3) both of the
previous effects result in a stronger Rayleigh scattering feature at a
fixed $N(H^o)$ and $U(H)$.  These differences are illustrated in
Figure~2b, comparing the two dashed lines.

The effective albedo of a matter-bounded cloud is represented by the
dot-dashed line in Figure~2b ($n(H) = 10^{11}~\rm{cm}^{-3}$, $U(H) =
10$, total hydrogen column density $N(H) = 10^{23}~\rm{cm}^{-2}$). The
effective albedo is grey as expected, and for this cloud this
characteristic extends between about 500~eV and 200~$\mu$m. At higher
energies metal opacities alter the effective albedo, and at wavelengths
longer than 200~$\mu$m the free-free opacity does so. If the column
density were larger at this ionization parameter, the greyness limits
at both ends would move toward the UV--optical regime. The free-free
opacity will be larger for larger gas densities. Under most conditions,
however, the $A_{eff}$(800--912~\AA\/) of a matter-bounded cloud
will remain grey with increasing total column density at constant
$U(H)$ until the optical depth at 912~\AA\/ approaches unity ($N(H^o)
\sim 10^{17.2}~\rm{cm}^{-2}$); the  $A_{eff}$(912~\AA\/--1~$\mu\/$m)
will remain grey until a hydrogen ionization front forms ($N(H) \gtsim
10^{23.1}~\rm{cm}^{-2} \times U(H)$).

Finally, except from clouds with small $N(H^o)$, the Rayleigh
scattering feature does not have a symmetric profile. With increasing
$N(H^o)$, the feature broadens, but the short wavelength side of the
profile is cut off by the ionization of \ci\/, excited state \oi\/, and
then ground state hydrogen and neutral oxygen at 912~\AA\/. For
$N(H^o)$ approaching $10^{24}$~cm$^{-2}$, the ionization of excited
state \caii\/ (near 1260~\AA\/), ground state \siI\/, \mgi\/, and
\fei\/ (1500~\AA\/ -- 1650~\AA\/) reduces the albedo on the long
wavelength side of the Rayleigh scattering profile. These opacity
features are labeled in Figure~2a.

\subsection{Reflection from Dusty, High Column Density Gas}

For purposes of comparison, we have plotted in Figure~2b (solid line)
the $A_{eff}(\lambda)$ from a high column density ($\log N(H) \approx
25$), dusty slab. The gas and grain abundances were assumed to
approximate that of the Orion nebula (Baldwin et al.\ 1996). The grains
used here tend to be larger and thus more grey (the ratio of total to
selective extinction is $R_V = 5.5$) in their optical properties than
those of the diffuse Galactic ISM. The larger grain size distribution
might be more appropriate for the AGN environment if we are observing
gas just exposed from a molecular cloud. The hydrogen particle density
was taken to be $10^{8}$~cm$^{-3}$; the flux in hydrogen ionizing
photons was set to $10^{17.25}$~s$^{-1}$cm$^{-2}$, just at the
threshold for the sublimation of silicate grains. The corresponding
ionization parameter was $\log U(H) = -1.23$. This was meant to be
grossly representative of the obscuring dusty torus (e.g., Pier \& Voit
1995), though the effective albedo of dusty gas depends little upon
these latter details. The effective albedo is roughly flat over much of
the optical -- ultraviolet, and reaches values no larger than about
0.2. The dust albedo diminishes substantially below 2500~\AA\/ and
above 1 micron. This is because, in our treatment, we discount forward
scattering so the grains become good absorbers at long and short
wavelengths. Other grain compositions and size distributions will have
albedos whose spectral shapes may differ, but whose overall amplitudes
will not. It is notable that the dusty gas cloud of low Thomson optical
depth (0.051) has a non-zero effective albedo at the Lyman limit
(0.032), reaches a maximum of 0.052 near 1.54 Rydbergs and declines at
higher energies.  The \Lya\/ Rayleigh scattering component is usually
very weak or absent because of the large grain opacity. While the dusty
cloud albedo never gets very high, it can compete with and even
dominate over the BLR gas albedo longward of 2000~\AA\/ if the ratio of
continuum source covering fractions (and so visibility) favors the
dusty gas. This is discussed in the next section.

\section{Discussion}

\subsection{The Integrated BLR Albedo}

The integrated BLR albedo is to first order a sum over each cloud's
$A_{eff}(\lambda) \times f_c$, where $f_c$ is that cloud's covering
fraction as seen by the continuum source. Of course, only those clouds
whose illuminated sides are observable will contribute to the observed
integrated BLR albedo. We can expect that this corresponds to roughly
1/2 of the integrated cloud covering fraction, on average.  We also
expect that clouds with a broad range in properties will contribute to
the integrated BLR albedo, rather than just the few types illustrated
here. The shapes of clouds, their distribution in space, and viewer
orientation will also play roles. We will not consider these
complications here.

It is apparent from Figure~2 that clouds with significant Thomson
optical depths, especially fully ionized ones, will contribute most to
the integrated BLR albedo at all wavelengths but in the vicinity of
\Lya\/.  In the spectral region $1200 \pm 200$~\AA\/ a wide variety of
ionization-bounded clouds of lower ionization parameter will contribute
most to the integrated albedo.  Next in importance to the albedo to the
Rayleigh scattering component in ionization-bounded clouds are those
spectral regions just to the long wavelength sides of the various
hydrogen recombination continuum edges, where the photoelectric
absorption and diffuse emission are at their minima.  Thus continuum
photons scattering off broad line clouds could alter the shapes of the
spectral features near the Paschen, Balmer, and Lyman jumps, reducing
the contrast at the continuum head.  The effect would be largest for a
population of high column density clouds with significant Thompson
scattering optical depths. For the model represented by the
light-dashed line in Figure~2b, the scattered continuum photons
represent 73\% of the total light emitted at 3649~\AA\/ from the
illuminated face of the cloud. However, unless the Thomson optical
depth is significant, the diffuse emission from most clouds will
dominate the reflected incident continuum, except in the vicinity of
\Lya\/.  While the contrast of either the diffuse emission or the
scattered continuum to the central continuum flux in our direction
would be expected to depend on the observer orientation and the spatial
distribution of clouds, to first order their ratio $F_{scat}/F_{diff}$
integrated over the BLR would remain approximately invariant to these
factors, since contributions to both of these quantities are emitted by
the same entities. However, $F_{diff}$ can also be emitted from the
backsides of clouds whose front sides are unviewable, and exact
radiative transfer through non-slab clouds could also alter the
predicted ratio.

The putative obscuring dusty torus likely covers a substantial portion
of the sky as seen from the central continuum source and BLR, and thus
could contribute significantly to the nuclear reflection spectrum in
AGN even when the central continuum source is visible. The ratio of
torus' covering fraction to that of the broad line clouds likely ranges
from roughly 1 -- 5 (e.g., Pier \& Krolik 1993), which means that the
integrated albedo of a molecular torus could be competitive with or
dominate over the BLR gas albedo longward of $\sim$~2000~\AA\/, though
the ratio of their {\em observable} emitting--reflecting surfaces is
not likely to be simply their ratio of covering fractions.  In addition
to the smooth incident continuum, the torus would also reflect the
diffuse emission from the broad line clouds. However, longward of 1
micron the dust re-emission will overwhelm the reflected light from the
continuum/BLR.

In summary, the integrated effective albedo over much of the
UV--optical for the combined BLR$+$torus mirrors should be a few to
several percent, or higher if the electron scattering albedo is
significant in the BLR. In the wavelength range 1130 -- 1330~\AA\/, the
Rayleigh scattering feature will likely dominate the observable
integrated effective albedo that could be as much as a few tens of
percent.

\subsection{Rayleigh Scattering Contribution to the Observed \Lya\/
Emission}

The Rayleigh scattering \Lya\/ feature is the most significant source
of albedo from ionization-bounded broad line clouds. From variability
studies (Clavel et al.\ 1991; Peterson et al.\ 1991; Korista et
al.\ 1995) we infer the existence of ionization-bounded clouds of high
column densities ($\log N(H) \gtsim 10^{23}$~cm$^{-2}$). Many of these
must have substantial $N(H^o)$. Thus, if the UV--optical albedo of
broad line clouds is important anywhere in the spectrum, it should be
in the vicinity of \Lya\/. Figure~2 shows features which range in width
from tens to hundreds of \AA\/ngstr\"oms, much like the observed broad
emission line profiles of \Lya\/. Might some fraction of the observed
equivalent width ($W_{\lambda}$) of \Lya\/ be due to continuum Rayleigh
scattering off the neutral portions of the broad line clouds?

In Figure~3 we plot the continuum normalized mean \Lya\/ profile of
NGC~5548, a Seyfert~1 galaxy, from the \HST\/ monitoring campaign of
1993 (Korista et al.\ 1995).  Also plotted is the function $F_{\lambda}
= 1 + A_{eff}(\lambda) \times 0.5f_c$, normalized at 1000~\AA\/ for 3
cloud covering fractions (0.1, 0.25, 0.5), for clouds with $\log U(H) =
-2$, $\log N(H^o) = 23$ (dotted lines). The factor of 0.5 in front of
the covering fraction assumes that this is the fraction of clouds whose
front faces are observable on average. The observable equivalent width
of the \Lya\/ Rayleigh scattering feature for clouds of these
parameters is $W_{\lambda\/1200}\rm{(Rayleigh)} \approx
100$~\AA\/~$\times 0.5f_c$, as measured between 1150~\AA\/ and
1300~\AA\/, over which one might attempt to measure the strength of
\Lya\/. For the effective albedo of this type of cloud, a maximum
covering fraction of $\ltsim$~0.5 is allowed, corresponding to an
observable $W_{\lambda\/1200}\rm{(Rayleigh)} \ltsim 25$~\AA\/. This
feature is very broad (FWHM$ \approx 23000$~km/s), and one of this
strength may account for the extremely broad wings observed in \Lya\/
(e.g., Zheng 1992) and the ``shelf'' upon which \oi\/ $\lambda$1304 and
\cii\/ $\lambda$1335 appear to sit.

Clouds with smaller $N(H^o)$ will have narrower Rayleigh scattering
profiles and smaller equivalent widths at a given covering fraction.
For comparison, the Rayleigh scattering profile of a cloud with $\log
U(H) = -2$, $\log N(H^o) = 22$ for $0.5f_c = 0.25$ is plotted as a
dashed line.  Its observable equivalent width is
$W_{\lambda\/1200}\rm{(Rayleigh)} \approx 40$~\AA\/~$\times 0.5f_c$,
while its FWHM is about 8800 km/s.  Clouds with $\log N(H^o) \ltsim 22$
will not contribute significantly to the measured {\em line} emission
at \Lya\/. Likewise is true for clouds with neutral hydrogen column
densities greatly in excess of $\rm{10^{23}~cm^{-2}}$, since the
reflection profile will be too broad to measure as an emission
line feature.

Since the BLR is composed of clouds with a variety of $U(H)$ and
neutral column densities, and since a scattered light spectrum is
always dependent upon the geometry, an upper limit to the contribution
of Rayleigh scattering to the measured \Lya\/ emission line strength is
not simple to predict, though a 10~\AA\/ equivalent width contribution
is a reasonable expectation. A significant contribution would
exacerbate the well-known \Lya\//\Hbeta\/ problem, described recently
in Netzer et al.\ (1995).

\subsection{Reflection of Optical--UV Photons from Thick, ``Neutral''
Media}

It should be obvious from Figure~2 that fully ionized gas of large
column densities, whether broad line clouds or atmospheres of accretion
disks, are not the only ``mirrors'' available in the vicinity of the
central engine of an AGN. Any source of significant $N(H^o)$ will
reflect continuum and line photons over many tens or hundreds of
\AA\/ngstr\"oms near 1200~\AA\/. Perhaps the broad line clouds are not
the most likely candidates for the very large neutral column density
clouds ($> 10^{24}$~cm$^{-2}$), but an accretion disk might be. As long
as the cool, relatively neutral portions of the accretion disk do not
contain dust and are not encased in an optically thick corona, then
Rayleigh scattering would be expected to contribute significantly to
the disk albedo in the UV. A glance at the heavy solid line in
Figure~2a shows that this Rayleigh scattering mirror could be 30\% to
nearly 100\% reflective over a $\sim$~700~\AA\/ span. Inwardly directed
line photons which hit this gas would be reflected; most of the UV
emission lines would be affected in this way to some degree (\ovi\/
$\lambda$1034 and \mgii\/ $\lambda$2800 would be least affected).
\Lya\/ is expected to be affected the most, since it is beamed fully
inward from ionization-bounded clouds (Ferland \& Netzer 1979), and the
effective disk albedo would be near unity near 1200~\AA\/. For lines
emitted near 1200~\AA\/, such as \nv\/ $\lambda$1240, this effect is
expected to be important even within broad line clouds of lower neutral
hydrogen column densities (see Figure~2a).

If this highly reflective Rayleigh mirror exists, it would not be
expected to lie in the inner portions of the accretion disk, where the
gas temperature is $\gtsim 10^5$~K and the neutral opacities are lower;
electron scattering will dominate the albedo under these conditions. In
addition to reflecting the inwardly streaming line photons, this warmer
electron scattering mirror would broaden the line profile (line photons
would retain their identity with the line as long as the temperature is
less than about $10^6$~K).  Significant line and/or continuum emission
scattered off the disk, whether by Rayleigh or electron scattering,
could have an impact on the observed emission line strengths, profiles,
and line-continuum reverberation. For instance, might reflection off
the disk symmetrize the \Lya\/ emission line profile?  Figure~4
illustrates a possible scenario whereby radiation emitted by the
illuminated face of a broad line cloud is redirected toward the
observer via reflection off an accretion disk.  Doppler shifts for
clouds moving toward or away from the reflecting disk would have the
opposite sign in reflection; Doppler motions parallel to the plane of
the disk would remain approximately invariant to reflection.  If the
net cloud motion were away from the disk, and toward the observer, the
portion of \Lya\/ that would be visible to the observer would have a
net blueshift (clouds on the far side of the disk are blocked from
view), in the absence of reflection from the disk. However, the
inwardly beamed \Lya\/ photons would be efficiently reflected from the
disk and redirected toward the observer with the opposite Doppler
shift.

Reflection off an accretion disk could solve a long-standing
conundrum.  Given the strength of the observed Balmer continuum, a
Lyman jump in emission is also expected.  However, one has never been
seen. The lack of a strong feature near 912~\AA\/, whether in
absorption (from the accretion disk?) or emission (from the broad line
clouds and/or accretion disk), has been a puzzle (Carswell \& Ferland
1988; Antonucci et al.\ 1996). One way to destroy a Lyman jump in
emission from broad line clouds is to reflect it off a source of
neutral hydrogen, perhaps the outer accretion disk, before we observe
it.

\subsection{Scattered Light in Broad Absorption Line Troughs}

One of the motivations of this work was that recent high-quality
spectropolarimetric observations of broad absorption line (BAL) QSOs
have shown that the BAL troughs are filled substantially with scattered
light (Glenn, Schmidt, \& Foltz 1994; Cohen et al.\ 1995; Goodrich \&
Miller 1995; Hines \& Wills 1995; Goodrich 1996). The resonance line
troughs in BAL QSOs are observed to be more polarized than the adjacent
continuum.  Investigators have proposed several scenarios, one of which
suggests that we observe two continuum sources. The primary continuum
is of little or no polarization, and is seen directly through the BAL
gas.  The secondary continuum results from the scattering of primary
continuum photons off some extended geometry into our line of sight
which takes a different path through the BAL resonance scattering
region (Figure~4 in Cohen et al.\ 1995). The total continuum is the sum
of the two and we see the strongly polarized light ($P \approx 10\%$)
only at the bottoms of the troughs where much of the primary,
unpolarized, continuum has been resonance-scattered from the line of
sight. Some unknown fraction of the scattered light at the bottoms of
the troughs is certain to be that from the scattering of the continuum
by the resonance lines which form the troughs (Hamann, Korista, \&
Morris 1993; Hamann \& Korista 1996; Lee \& Blandford 1996), and
possibly from dust or electron scattering within or interior to the BAL
flow. But assuming the ``extended secondary continuum'' scenario to be
correct, could some portion of the BLR contribute to its origin?

A significant fraction of those clouds emitting the broad emission
lines must be covered by the BAL outflow (Turnshek et al.\ 1988).  This
is clear from the weakness of \Lya\/ emission in many BAL QSOs due to
scattering by the N$^{+4}$ ions in the BAL outflow, and also from the
observed \civ\/ BAL profiles at lower outflow velocities.  However,
some fraction of the BLR might lie along sight-lines of lower or zero
optical depth through the BAL region and act as an extended continuum
source. Is there a spectral signature in the observed scattered
(polarized) light that we may identify as emerging from the BLR?

Figure~2 illustrates a wide variety of $A_{eff}(\lambda)$ expected from
BLR clouds. The albedo of BLR clouds in the UV results from some
combination of Rayleigh plus electron scattering that is diminished by
various atomic opacity sources. The wavelength dependence of the
reflectivity of the BLR mirror could be as featureless as pure electron
scattering or as ``blue'' as Rayleigh scattering, and is likely
somewhere in between. A Keck spectropolarimetric observation of the
high redshift BAL QSO, 0105$-$265 (Cohen et al.\ 1995), shows that the
polarized light does not diminish or disappear abruptly at wavelengths
shortward of 1100~\AA\/, 1064~\AA\/, and 912~\AA\/ as would be expected
for the case of broad line clouds with a significant $N(H^o)$. These
types of clouds cannot dominate this mirror's composition. The most
prominent sign of reflection by an ionization-bounded cloud of
significant column density is the Rayleigh scattering feature. However,
this has the misfortune of lying near the unpolarized broad \Lya\/
emission line, and is also susceptible to resonance line scattering by
the the \nv\/ ion in the outflowing gas. Determination of the origin(s)
of this scattered light will require disentangling the various possible
sources of scattered light from each other and from sources that dilute
the polarized light.

\section{Summary}

We have explored the effective UV--optical albedos of a variety of
possible broad emission line clouds, as well as their effects on the
observed spectra of AGN. The effective albedo results from an interplay
between the scattering and absorption opacities. In moderately ionized
ionization-bounded clouds the most important source of albedo is that
due to Rayleigh scattering of continuum photons off the extreme damping
wings of \Lya\/. The photons resulting from this scattering mechanism
may contribute significantly to the \Lya\/ emission line, especially in
the very broad wings. Broad line clouds with $N(H^o) \gtsim
10^{21}$~cm$^{-2}$ will reflect an increasingly significant fraction of
the \nv\/ $\lambda$1240 photons that stream toward the neutral portion
of the cloud. Other emission lines may be similarly affected to
differing degrees. Inclusion of this effect can alter predicted
emission line strengths and profiles because the line beaming function
becomes more anisotropic. The effective albedo from a significant
population of highly ionized, ionization-bounded clouds arises from
Thomson scattering, and diminished by the neutral hydrogen opacity. The
interesting feature here is that the effective albedo tends to be
maximum just {\em longward} of the continuum edges.  Thus a significant
albedo from such clouds could alter the shape of the spectrum near the
Balmer jump, for example.  Fully ionized clouds will have nearly grey,
Thomson scattering, UV--optical albedos. A thick, dusty cloud will
contribute a fairly grey near-UV to optical albedo. In the absence of a
strong Thomson scattering mirror, the dusty torus could dominate the
albedo from the nuclear regions longward of $\sim$2000~\AA\/ due to the
torus' expected large covering fraction.  The accretion disk may act as
electron and Rayleigh scattering mirrors of broad emission line photons
as well as central continuum photons, but its impact on the observed
UV--optical spectrum of quasars awaits a fuller understanding of the
accretion phenomenon.  Finally, ionization-bounded BLR clouds may be a
contributor to the mirror proposed to explain the polarized BAL
troughs.

\acknowledgements
This work was supported by NASA (NAG-3223) and NSF AST 96-17083. 
We thank Bob Goodrich and Patrick Ogle for helpful discussions, and
we are grateful to an anonymous referee for his or her constructive
comments.
%
%
%

%
\newpage
%
%
%
\begin{center}
{\bf Figure Captions}
\end{center}

\noindent
Fig. 1.--- {A cartoon illustrating the source of continuum photons at
left (star icon) and a broad emission line cloud at right. The incident
and reflected continua and the diffuse emission are labeled. As defined
here, the reflected continuum does not include diffuse emission.}

\noindent
Fig. 2a.--- {The optical -- ultraviolet effective continuum albedo as a
function of $\log \lambda$ (\AA\/) for clouds with a range in $N(H^o)$
and $U(H) = 0.01$, as labeled in the plot. Significant gas absorption
opacity features are labeled. The bold curve includes the effects of introducing a turbulent velocity field with $\sigma\/ = 100$ \kms\/.}

\noindent
Fig. 2b.--- {The optical -- ultraviolet effective continuum albedo as a
function of $\log \lambda$ (\AA\/) for a variety of cloud parameters,
as labeled in the plot. The dot-dashed represents the albedo of a
matter-bounded cloud whose {\em total} column density is
$10^{23}~\rm{cm}^{-2}$. The solid line shows the effective albedo of a
dusty, high column density cloud, $N(H^o) \approx
10^{25}~\rm{cm}^{-2}$.}

\noindent 
Fig. 3.--- {Solid line: continuum normalized mean 1993 HST spectrum of
NGC~5548. Dotted Lines: the function $1 + A_{eff}(\lambda) \times
0.5f_c$, normalized at 1000~\AA\/ for 3 cloud covering fractions (0.1,
0.25, 0.5). The effective albedo is from a cloud with $\log U(H) = -2$
and $\log N(H^o) = 23$. Dashed line: the same function from a cloud
with $\log N(H^o) = 22$ and $0.5f_c = 0.25$. Geocoronal \Lya\/
contaminates the blue mid-wing of the \Lya\/ emission profile of
NGC~5548.  Note that both axes are on a $\log_{10}$ scale.}

\noindent 
Fig. 4.--- {A cartoon illustrating the effect of an accretion disk
reflecting inwardly directed emission from a broad line cloud. The
central ionizing source and the accretion disk are denoted by the star
icon and shaded ellipse. Inwardly beamed radiation from the cloud,
otherwise invisible to the observer, is reflected toward the smiling
observer. Emission line clouds on the other side of the disk from the
observer are invisible to the observer.}


\begin{references}

\reference{} Antonucci, R.R.J., Geller, R., Goodrich, R.W., \& Miller,
J.S.\ 1996, \apj, 472, 502

\reference{} Antonucci, R.R.J., \& Miller, J.S.\ 1985, \apj, 297, 621

\reference{} Baldwin, J.A., Ferland, G.J., Korista, K.T., \& Verner,
D.A.\ 1995, \apj, 455, L119

\reference{} Baldwin, J.A., et al.\ 1996, \apj, 468, L115

\reference{} Carswell, R.F., \& Ferland, G.J.\ 1988, \mnras, 235, 1121

\reference{} Clavel, J., et al.\ 1991, \apj, 366, 64

\reference{} Cohen, M.H., Ogle, P.M., Tran, H.D., Vermeulen, R.C.,
Miller, J.S., Goodrich, R.W., \& Martel, A.R.\ 1995, \apj, 448, L77

\reference{} Ferland, G.J., 1996, in {\sc Hazy}, Univ.\ Kentucky
Dept.\ Phys.\ \& Astron.\ Internal Rep.

\reference{} Ferland, G.J., \& Netzer, H.\ 1979, \apj, 229, 274



\reference{} Gavrila, M.\ 1967, Physical Review, 163, 147

\reference{} Glenn, J., Schmidt, G.D., \& Foltz, C.B.\ 1994, \apj, 434,
L47

\reference{} Goodrich, R.W.\ 1997, \apj, 474, 606

\reference{} Goodrich, R.W., \& Miller, J.S.\ 1994, \apj, 434, 82

\reference{} Goodrich, R.W., \& Miller, J.S.\ 1995, \apj, 448, L73

\reference{} Grevesse, N., \& Anders, E.\ 1989, in  AIP
Conf.\ Proc.\ 183, Cosmic Abundances of Matter, ed.\ C.J.\ Waddington
(New York: AIP), p.\ 1

\reference{} Grevesse, N., \& Noels, A., 1993, in Origin and Evolution
of the Elements, ed.\ N.\ Prantzos, E.\ Vangioni-Flam, \& M.\ Casse
(Cambridge Univ.\ Press), p.\ 15

\reference{} Hamann, F., Korista, K.T., \& Morris, S.L.\ 1993, \apj,
415, 541

\reference{} Hamann, F., \& Korista, K.T.\ 1996, \apj, 464, 158

\reference{} Hines, D.C., \& Wills, B.J.\ 1995, \apj, 448, L69

\reference{} Koratkar, A., Antonucci, R.R.J., Goodrich, R.W., Bushouse,
H., \& Kinney, A.L., 1995, \apj, 450, 501

\reference{} Korista, K.T., et al.\ 1995, \apjs, 97, 285

\reference{} Krolik, J.H., Madau, P., \& \.{Z}ycki, P.T.\ 1994, \apj,
420, L57

\reference{} Lee, H.W., \& Blandford, R.D.\ 1996, \apj, in press

\reference{} Lightman, A.P., \& White, T.R.\ 1988, \apj, 335, 57

\reference{} Miller, J.S., \& Antonucci, R.R.J.\ 1983, \apj, 271, L7

\reference{} Miller, J.S., \& Goodrich, R.W.\ 1990, \apj, 355, 456

\reference{} Mihalas, D.\ 1978, Stellar Atmospheres, 2nd Edition (San
Francisco: W.H.\ Freeman)

\reference{} Nandra, K., \& George, I.M.\ 1994, \mnras, 267, 974

\reference{} Netzer, H., Brotherton, M.S., Wills, B.J., Han, M., Wills,
D., Baldwin, J.A., Ferland, G.J., \& Brown, I.W.A.\ 1995, \apj, 448, 27


\reference{} Peterson, B.M., et al.\ 1991, \apj, 368, 119

\reference{} Pier, E.A., \& Krolik, J.H.\ 1993, \apj, 418, 673

\reference{} Pier, E.A., \& Voit, G.M.\ 1995, \apj, 450, 628

\reference{} Poutanen, J., Sikora, M., Begelman, M.C., \& Magdziarz,
P.\ 1996, \apj, 465, L107

\reference{} Ross, R.R., \& Fabian, A.C.\ 1993, \mnras, 261, 74

\reference{} Tran, H.D., Miller, J.S., \& Kay, L.E.\ 1992, \apj, 397,
452

\reference{} Turnshek, D.A., Foltz, C.B., Grillmair, C.J., \& Weymann,
R.J.\ 1988, \apj, 325, 651

\reference{} Walter, R., Orr, A., Courvoisier, T.J.-L., Fink, H.H.,
Makino, F., Otani, C., \& Wamsteker, W.\ 1994, \aap, 285, 119

\reference{} Zheng, W.\ 1992, \apj, 385, 127

\reference{} \.{Z}ycki, P.T., Krolik, J.H., Zdziarski, A., \& Kallman,
T.R.\ 1994, \apj, 437, 597


\end{references}
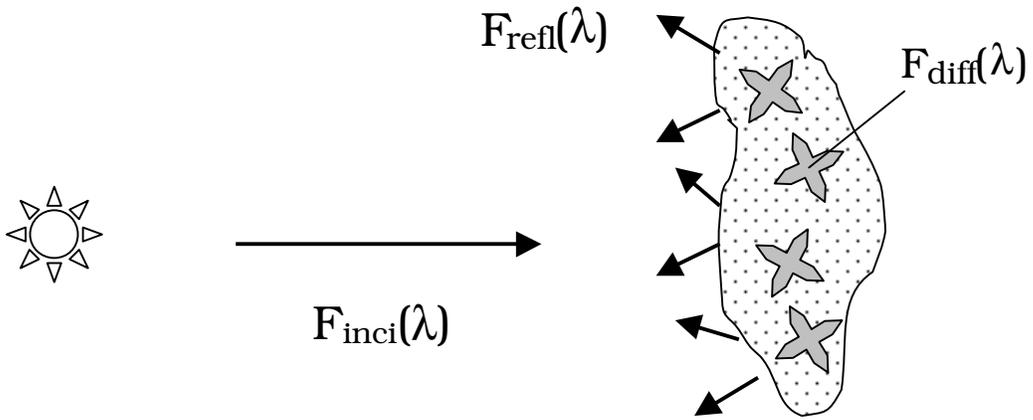
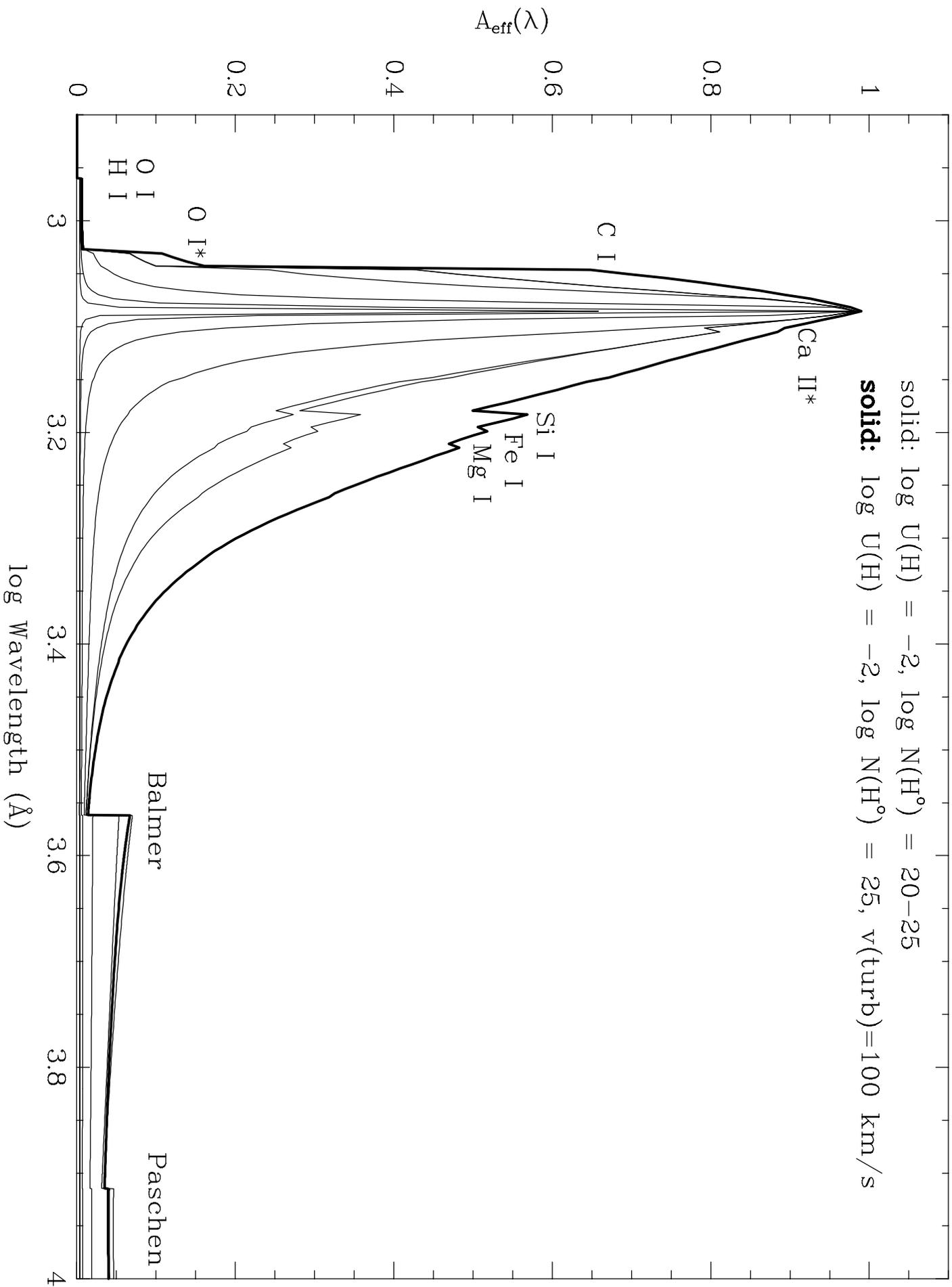
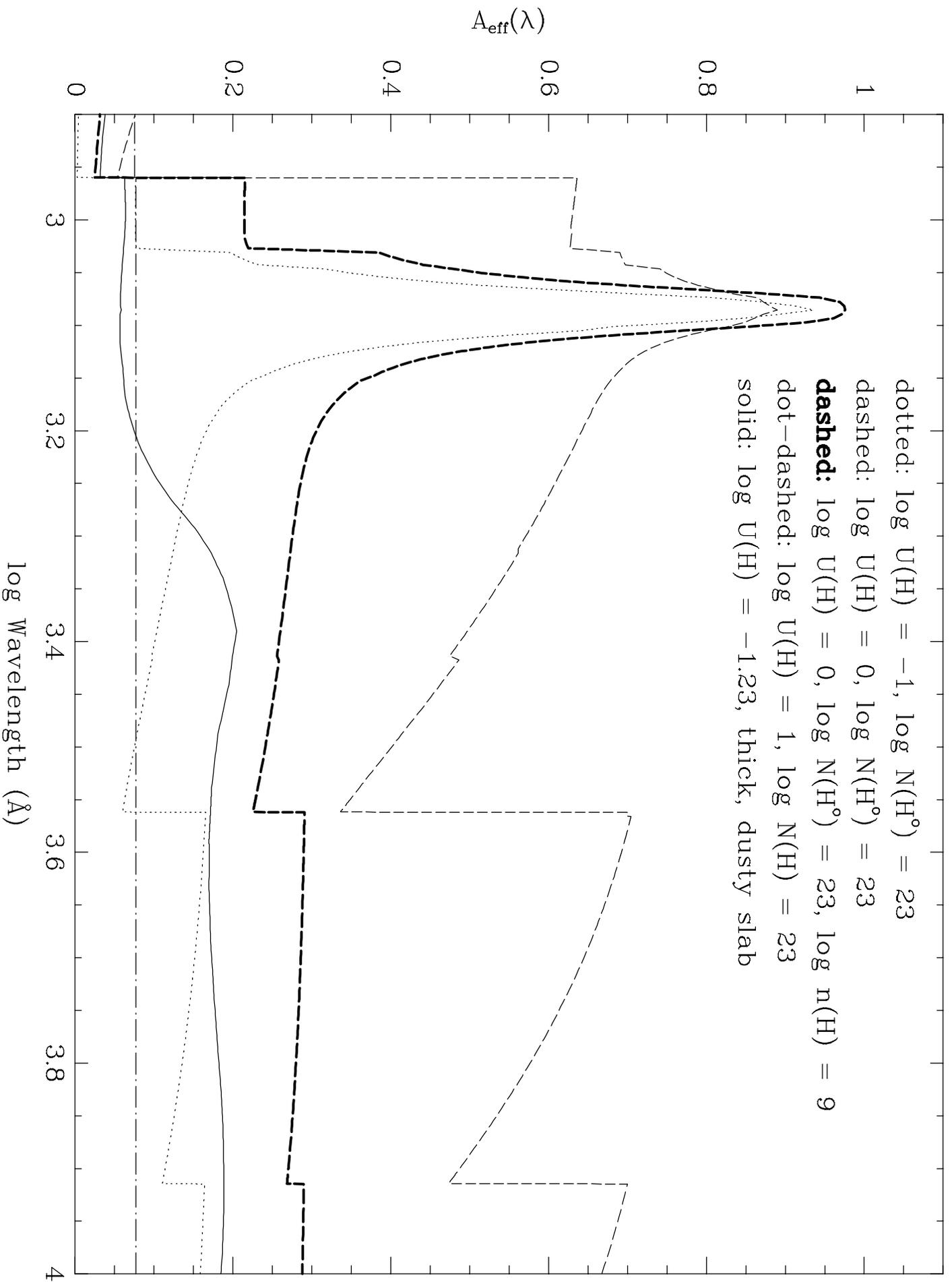
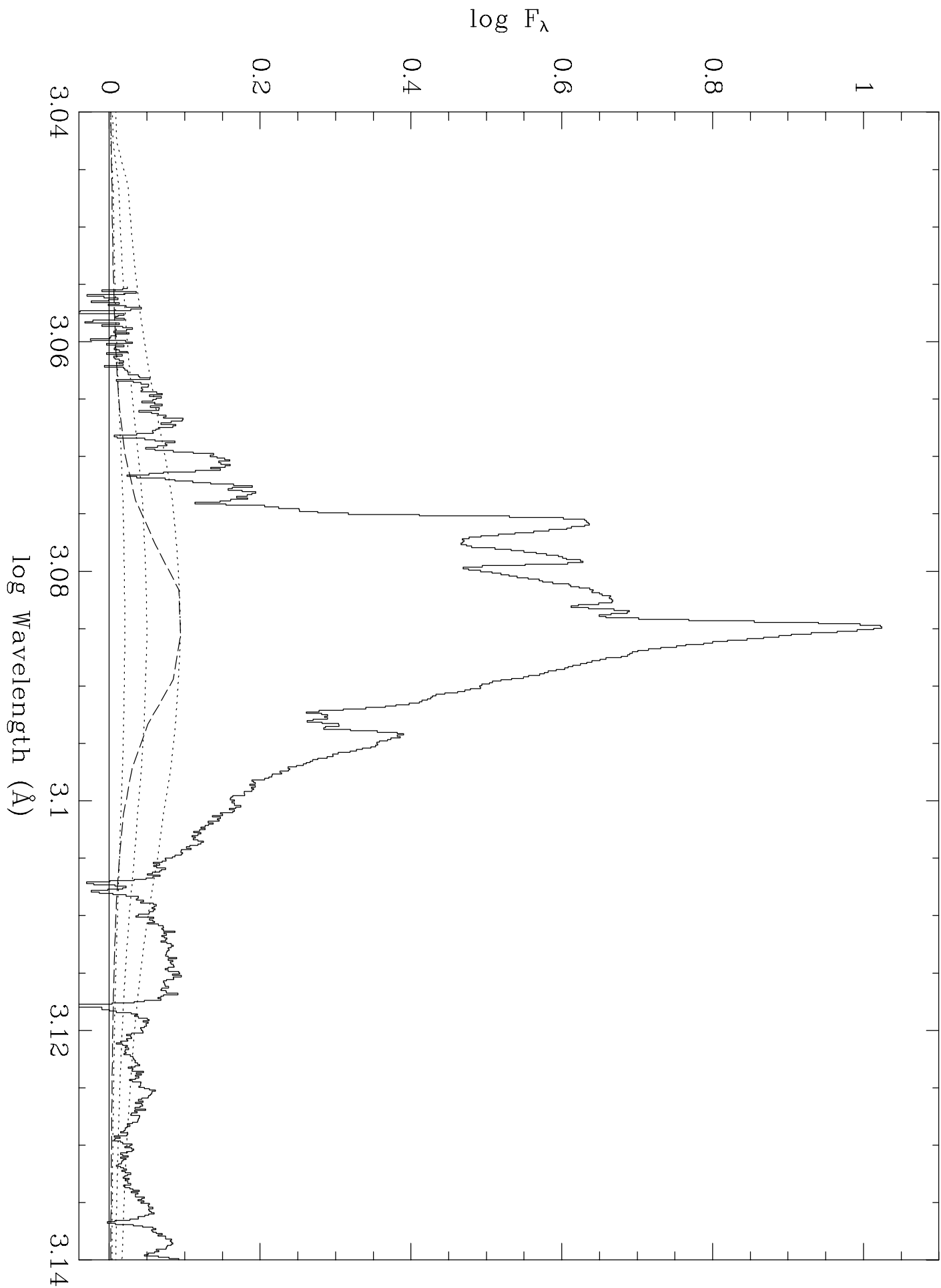
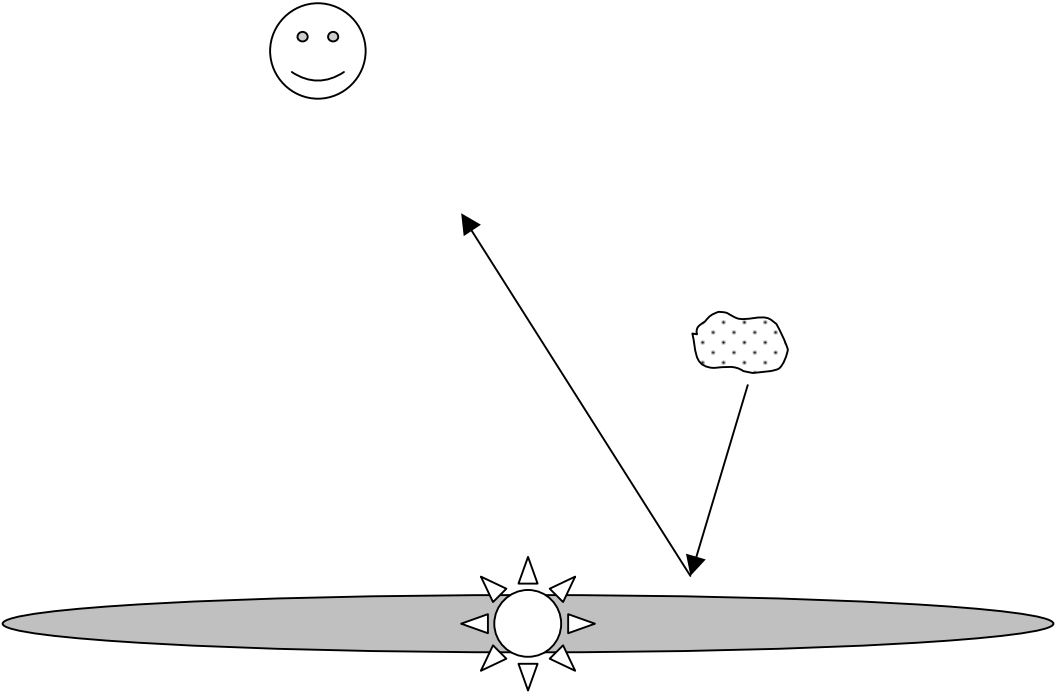
\end{document}